\documentclass[notitlepage, twoside,english,nofootinbib,preprintnumbers,aps,11pt,showpacs, superscriptaddress, groupedaddress]{revtex4-1} \usepackage[a4paper, margin=1.5in]{geometry} \usepackage{amssymb} \usepackage{commath}
\usepackage{amsmath}
\usepackage{url}
\usepackage{float}
 \usepackage[utf8]{inputenc} \usepackage[T1]{fontenc} \usepackage{lmodern}
\usepackage{braket}
\usepackage{graphicx}
\usepackage{pictexwd,dcpic}
\usepackage{braket}
\usepackage{atbegshi,picture, afterpage}
\usepackage{lipsum}
\usepackage{physics}
\usepackage{appendix}
\usepackage{bbm}
\usepackage{hyperref}
\usepackage{color}

\newcommand{\be}{\begin{equation}}
\newcommand{\ee}{\end{equation}}
\newcommand{\ba}{\begin{align}}
\newcommand{\ea}{\end{align}}
\begin{document}
\title{Energy of gravitational radiation in the de Sitter universe at the scri and at a horizon}
\author{Maciej Kolanowski}
\email[]{Maciej.Kolanowski@fuw.edu.pl}
 \address{\vspace{6pt} Institute of Theoretical Physics, Faculty of
   Physics, University of Warsaw, Pasteura 5, 02-093 Warsaw, Poland}
  \author{Jerzy Lewandowski}
\email[]{Jerzy.Lewandowski@fuw.edu.pl}
\address{\vspace{6pt} Institute of Theoretical Physics, Faculty of
  Physics, University of Warsaw, Pasteura 5, 02-093 Warsaw, Poland}
  \date{\today}
\begin{abstract} \noindent
 In this note we investigate outcomes of a symplectic formula for the gravitational waves charges in the general relativity linearized around the de Sitter spacetime. We derive their explicit form at {\it scri} in the Bondi frame, compare with the connected Noether expression and analyze their gauge dependence which allows us to fix unambiguously boundary terms. We also discuss minimal requirements needed to impose on initial data to have finite values of charges. Furthermore, we analyze transformation laws of the energy upon the action of the de Sitter group and discuss its physical interpretation. Finally, we calculate its flux through a cosmological horizon instead of {\it scri}. We show that in the limit $\Lambda \to 0$, one recovers Trautman--Bondi formula strengthening recent proposal that one should choose a~null surface as a more natural boundary for the astrophysical systems in the presence of the cosmological constant. 
\end{abstract}
\maketitle
\section{Introduction} \noindent 
Gravitational waves (GW) in the presence of the positive cosmological constant $\Lambda$ attracted a lot of attentions in the last few years -- different notions of energy were defined (e.g. \cite{kastor2002positive, Penrose:2011zza, Penrose:2011zza, Ashtekar:2015lla, Szabados:2015wqa, Chrusciel:2020rlz}), different boundary conditions (or gauges) were imposed or reject (e.g. \cite{Ashtekar:2014zfa, He:2015wfa, Poole:2018koa, Compere:2019bua}) and finally applied in physical situations (e.g. \cite{Ashtekar:2015lxa, Bishop:2015kay, Date:2015kma}). The reasons for such a state of affairs are quite clear -- there is no denying nowadays that both GWs are real \cite{Abbott:2016blz} and $\Lambda > 0$ \cite{riess1998observational}. Unfortunately, our understanding of those two issues {\it combined} is still very preliminary. All our intuitions (like Bondi news tensor or a universal structure at infinity) stem from the asymptotically flat case which is qualitatively different. We do not wish to propose yet another framework  or definition in this paper but rather provide some additional insight into the symplectic approach of \cite{Ashtekar:2015lla}. First of all, we will compare it with the more recent proposal based upon Noether currents \cite{Chrusciel:2016oux}. We investigate the differences (which lie in the boundary terms) finding that both of them have certain drawbacks and ambiguities which can be uniquely fixed by the requirement of a~gauge invariance. Later on, we will also discuss transformation properties of the $\Lambda$-Trautman--Bondi energy under de Sitter isometries and provide its physical interpretation. We finish by repeating our calculation at a cosmological horizon and analyzing the $\Lambda \to 0$ limit for the fields living in its neighbourhood.
\\
The rest of this paper is organised as follows. In the Sec. \ref{bondi} we calculate de Sitter charges in the Bondi frame using the symplectic form. Although the result is finite, there is a usual ambiguity regarding boundary terms. We compare it with the approach of \cite{Chrusciel:2016oux} which used Noether current as a starting point. 
Within the Sec. \ref{allowed} we rewrite obtained charges in terms of the initial data at $\mathcal{I}^+$. This new form will enable us to establish boundary conditions needed for their convergence, compare with the analogous formulas in a Poincar\'e patch \cite{Ashtekar:2015lla} and finally discuss gauge dependence of our results. In the Sec. \ref{komutatory} we discuss how the structure of the de Sitter group affects transformation properties of the GW's energy and how it can be realized and interpreted in different settings. In the Sec. \ref{horizon} we calculate flux of the energy of the linearized gravity through the cosmological horizon and discuss its $\Lambda \to 0$ limit. In the Sec. \ref{conclusions} we summarize our results. We relegate computational details to the Appendixes. 
\section{Canonical energy in the Bondi frame} \label{bondi}
\subsection{Preliminary notions} 
\noindent
It is natural to associate an energy with a timelike Killing vector. Unfortunately, since {\it scri} $\mathcal{I}^+$\footnote{the same applies to the asymptotic past $\mathcal{I}^-$. We will refer to the {\it scri} as $\mathcal{I}^+$ only for concreteness.} is spacelike and it is preserved by all de Sitter isometries, there is no globally timelike Killing vector but we still can choose one which is timelike in some part of the spacetime. Additionally, one should require that chosen generator reduces to $\partial_t$ upon the İnönü--Wigner contraction as $\Lambda \to 0$. De Sitter metric can be written in the Bondi coordinates as
\begin{equation}
    g = - \left(1 - \frac{\Lambda r^2}{3} \right) du^2 - 2dudr + r^2 \mathring{\gamma}_{AB} dx^A dx^B, \label{g}
\end{equation}
where $\mathring{\gamma}_{AB}$ is a standard metric on a unit sphere and
\begin{equation}
    \Lambda > 0.
\end{equation} It seems thus natural to choose
\begin{equation}
    T = \partial_u
\end{equation}
 as our 'time-translation' generator.  One can easily check that spatial translations generators\footnote{We call them that way because they generate $3$-dimensional abelian subgroup which act exactly as translations in the Poincar\'e patch.} 
\begin{equation}
    T_{(i)} = e^{\sqrt{\frac{\Lambda}{3}}u}
    \left(
    g_{i} \partial_u -  g_{i} \left(Hr + 1 \right) \partial_r - \left( \frac{1}{r} + H \right)\mathring{D}^A g_{i} \partial_A
    \right), \label{translacje} 
\end{equation}
where $g_i (x^A)$ are proportional to the $l=1$ spherical harmonics, commute with $T$ in the limit $\Lambda \to 0$ as required. \\
Throughout this paper we consider a perturbed metric tensor
$$g + h,$$ 
and assume the perturbation $h$ satisfies the linearised vacuum Einstein equations with the cosmological constant $\Lambda$ on the background  $g$. The perturbation $h$ is defined in the same domain as the coordinates $(u,r,\theta,\phi)$ and is subject to asymptotic conditions formulated in next subsection. The global issues are discussed in the Section 3.  \\
Let us also briefly mention frequently used notations. Hubble parameter  is given by $H = \sqrt{\frac{\Lambda}{3}}$. It is natural to consider 3-dimensional metric $\mathring{q}$ induced on {\it scri} $\mathcal{I}^+$. We use a conformal factor $\Omega = H^{-1} r^{-1}$ and so it reads
\begin{equation}
    \mathring{q} = du^2 + H^{-2} \mathring{\gamma}_{AB} dx^A dx^B.
\end{equation}
By $\mathring{D}_A$ we denote the covariant derivative of $\mathring{\gamma}$. \\
We shall also use the electric part of the Weyl tensor which is given by
\begin{equation}
    E_{ac} = C_{abcd} n^b n^d, 
\end{equation}
where $C_{abcd}$ is the (linearized) Weyl tensor of $g+h$ and $n$ is the vector normal to the surface $r = \textrm{const.}$ of length $-1$. One can check that $E_{ac}$ vanishes as $r \to \infty$ and so we introduce
\begin{equation}
    \mathcal{E}_{ac} = \lim_{r \to \infty} \Omega^{-1} E_{ac}
\end{equation}
Since the de Sitter is conformally flat, $\mathcal{E}$ is gauge invariant. To avoid baroque names, we will simply call $\mathcal{E}$ the electric part of the Weyl tensor.
\subsection{Noether charge} \noindent
Recently, GW's energy was  defined and calculated by using the  Noether charge  (of the linenarized theory) associated with the vector field $T$ in the Bondi frame \cite{Chrusciel:2020rlz} and some regularisation  procedure necessary to obtain a finite result.  We make the same assumptions about the perturbations $h$ of the de Sitter metric $g$, while our goal is application of the symplectic approach. For the completeness, we also remind the reader the  results obtained therein. 
Following \cite{friedrich1986}, metric perturbations $h$ had the following asymptotics as $r \to \infty$ ($x$ stands for the two variables $x^A$): 
\begin{align}
\begin{split}
    \frac{h_{AB}(r,u,x)}{r^2} &= \frac{h_{AB}^{(-1)} (u,x)}{r} + \frac{h_{AB}^{(-3)} (u,x)}{r^3} +... \\
     \frac{h_{Au} (r,u,x)}{r^2} &= {h}^{(0)}_{Au}(u,x) + \frac{\mathring{D}^B h_{AB}^{(-1)} (u,x)}{2r^2} + \frac{h^{(-3)}_{Au} (u,x)}{r^3} +... \label{bondi_exp}
\end{split}
\end{align}
Moreover, 
\begin{equation}\label{gauge}h_{ra} = 0,\quad {\rm and}\quad  {h}_{AB}\mathring{\gamma}^{AB}=0.\end{equation} 
In fact, the choice $h_{ur}$ can be only imposed at {\it scri}. However, in the {\it linearized} theory, $h_{ur,r} = 0$ and thus it is valid everywhere. The lack of the $h^{(-2)}_{AB}$ term assures that this expansion is smooth in $r^{-1}$ (there is no logarithmic dependence) \cite{Chrusciel:2016oux, Compere:2019bua} and in fact it vanishes on-shell as long as $\Lambda \neq 0$ and $\mathcal{I}^+$ is smooth. This is a little bit different version of the Bondi gauge than the radial one used in \cite{Ashtekar:2015ooa, Compere:2019bua} where the leading term is $h_{AB}$ rather than
\begin{equation}
    h_{uA} =: h_A
\end{equation}
The coefficient $h_{AB}^{(-1)}$ is determined by the $h_{A}^{(0)}$ through the following constraint:
\begin{equation}
    \mathring{D}_A h_{B}^{(0)} + \mathring{D}_B h_{A}^{(0)} - \mathring{D}^C  h_{C}^{(0)} \mathring{\gamma}_{AB} = - \frac{\Lambda}{3} h^{(-1)}_{AB}  \label{constraint}.
\end{equation}
Later on, the authors calculated the Noether energy stored in a null cone $C_u$
$$ u=\rm const. $$
and promoted the difference between two cones to be the radiated energy. Unfortunately, they obtained infinite result which needed some form of regularisation. (Let us notice that it was simply a matter of integration by parts and attaching boundary terms towards different regions). Their final result $E^{\rm N}(u)$ satisfies an analog of the Trautman--Bondi mass loss formula, namely \cite{Chrusciel:2020rlz}
\begin{align}
&E^{\rm N}(u_1)-E^{\rm N}(u_2)=\nonumber\\
&\frac{1}{32 \pi}\int_{u_1}^{u_2} du\int_0^\pi \sin \theta d \theta\int_0^{2\pi} d\phi \mathring{\gamma}^{AB} \left( \mathring{\gamma}^{CD} h^{(-1)}_{AC,u} h^{(-1)}_{BD,u} - 6 h^{(-3)}_{A} h^{(0)}_{B,u} 
    \right). \label{chrusciel}
\end{align}
The domain of the integration corresponds to the segment of the future {\it scri} $\mathcal{I}^+$ contained between the intersections with the cones $C_{u_1}$ and $C_{u_2}$, respectively. Technically, this is the limit as $r\to\infty$ of the segment $\Sigma_r (u_2,u_1)$ of the surface
\begin{equation}
    r = \textrm{const.}
\end{equation}
connecting the cones. A nice feature of this expression is that in the limit $\Lambda \to 0$  we have $\stackrel{(0)}{h}_{B} \to 0$ and thus the usual Trautman--Bondi formula \cite{Trautman:2016xic, Bondi:1962px, Sachs:1962wk} is recovered\footnote{One should understand this limiting procedure in the following way: there is a family of solutions to the linearized Einstein equations, parametrized by $\Lambda$, for which $\stackrel{(0)}{h}_{B} \to 0$ as $\Lambda \to 0$ and for such family energy has expected limit. It can be given e.g. by $\Lambda$-independent $\stackrel{(-1)}{h}_{AC}$ and then $\stackrel{(0)}{h}_{B}$ is an appropriate solution to the constraints \eqref{constraint} which can be consistently chosen to be proportional to $\Lambda$.}. 
\subsection{Symplectic formulation} \noindent
Having discussed the state-of-the-art, let us now compare with the symplectic approach.
Given a solution $g$ to  $\Lambda$-vacuum Einstein's equations, and two perturbations $h_1$ and $h_2$ that satisfy the linearised $\Lambda$-vacuum Einstein equations, the corresponding symplectic current is the vector field  $\omega^a(h_1,h_2)$ defined in spacetime by the following formulae
\begin{equation}
    \omega^a(h_1, h_2) = \frac{1}{16 \pi}  P^{abcdef} \left(
    h_{2\ bc} \nabla_d h_{1\ ef} - h_{1\ bc} \nabla_d h_{2\ ef}
    \right),\label{sympl}
\end{equation}
where
\begin{equation}
    P^{abcdef} = g^{ae} g^{fb} g^{cd} - \frac{1}{2} g^{ad} g^{be} g^{fc} - \frac{1}{2} g^{ab} g^{cd}g^{ef} - \frac{1}{2} g^{bc} g^{ae} g^{fd} + \frac{1}{2} g^{bc} g^{ad} g^{ef}. 
\end{equation}
The symplectic current is divergence free
$$ \nabla_a \omega^a(h_1, h_2)=0.$$
In the case at hand $g$ is the de Sitter solution (\ref{g}), $h_1$ is our perturbation $h$ and 
$$h_2={\cal L}_X h \label{X}$$
where $X$ is a Killing vector field of $g$. Notice that ${\cal L}_X h$ satisfies the linearised $\Lambda$-vacuum Einstein equations, since $h$ does, however, depending on $X$, it may or may not satisfy the gauge conditions (\ref{gauge}) . 

Suppose $X$ is the Killing vector $T=\partial_u$.  Then, the following integral along a~cone $C_u$ \begin{equation}  
E(u) = \frac{1}{2} \int_{C_u}\omega^a(h,{\cal L}_Th)\epsilon_{abcd}\frac{1}{3!}dx^bdx^cdx^d
\end{equation}
is the total energy of the perturbation $h$ contained in the  cone $C_u$.
The radiated energy  is the difference 
\begin{equation}
E(u_1)-E(u_2) =- \frac{1}{2}{\rm lim}_{r \to \infty}\int_{\Sigma_r(u_2,u_1)}\omega^a(h,{\cal L}_Th)\epsilon_{abcd}\frac{1}{3!}dx^bdx^cdx^d \label{roznica}
\end{equation}
Tedious yet straightforward calculation gives the following result:
 \begin{align}     E(u_1)-E(u_2) = \frac{1}{32 \pi}\int_{u_1}^{u_2} du\int_0^\pi \sin \theta d \theta\int_0^{2\pi} d\phi \nonumber\\
       \left(\mathring{\gamma}^{BE} \left(\mathring{\gamma}^{FC} h_{BC,u}^{(-1)} h_{EF,u}^{(-1)} - 6 h^{(-3)}_B h^{(0)}_{E,u}\right)  - \frac{1}{2}\mathring{\gamma}^{BE} \partial_u \left(\mathring{\gamma}^{FC} h_{BC}^{(-1)}h_{EF,u}^{(-1)}  - 6h^{(-3)}_B h^{(0)}_E\right)\right) \label{energy_flux}
\end{align}
which is equal to \eqref{chrusciel} modulo boundary terms. However, in comparison with the Noether current approach, there is one crucial difference -- integrand of \eqref{energy_flux} is automatically finite, no regularisation is needed because all a priori divergent terms cancel out {\it on-shell}. On the other hand those additional boundary terms are somehow troublesome because they spoil the $\Lambda \to 0$ limit. Obviously, the total emitted energy does not depend on them but if one would like to use or measure it, it is necessary to know not only the global quantity but also its distribution\footnote{Since $\mathcal{I}^+$ is spacelike, different values of $u$ coordinate in a sense correspond rather to different locations then times of arrival, in contrast with the asymptotically flat case.}. \\
For the future use, it is convenient to define the total radiated energy
\begin{equation}
    Q_T[h] = \frac{1}{32 \pi}\int_\mathbb{R} du\int_0^\pi \sin \theta d \theta\int_0^{2\pi} d\phi \nonumber\\
       \mathring{\gamma}^{BE} \left(\mathring{\gamma}^{FC} h_{BC,u}^{(-1)} h_{EF,u}^{(-1)} - 6 h^{(-3)}_B h^{(0)}_{E,u}\right) \label{energia_total}
\end{equation}
which a priori does not have to be finite and for convenience we already discarded boundary terms. We analyze this issue more thoroughly in the Sec. \ref{allowed}.
\subsection{Angular momentum}\noindent
We will now calculate angular momenta corresponding to the rotations around the line $r=0$ and thus well-adapted to the Bondi frame. Therefore, let  the Killing vector field $X$ in (\ref{X}) be
\begin{equation}
    S := \partial_\phi.
\end{equation}
The symplectic flux
\begin{equation}  
J(u) = -\frac{1}{2} \int_{C_u}\omega^a(h,{\cal L}_Sh)\epsilon_{abcd}\frac{1}{3!}dx^bdx^cdx^d
\end{equation}
is the  angular momentum of the perturbation $h$ contained in the  cone $C_u$. The total radiated angular momentum is
\begin{align}
\begin{split}
J(-\infty) - J(\infty) &=: \\
    Q_S[h] =\frac{1}{64\pi} \int_{\mathbb{R}} du \int_0^\pi \sin \theta d\theta \int_0^{2\pi}d\phi &\mathring{\gamma}^{BE} \mathring{\gamma}^{FC}\left( \mathcal{L}_S h^{(-1)}_{BC}  h^{(-1)}_{EF,u} -h^{(-1)}_{BC} \mathcal{L}_S h^{(-1)}_{EF,u} \right) \\
    &-6 \mathring{\gamma}^{AB} \left(\mathcal{L}_S h^{(-3)}_A h^{(0)}_B - \mathcal{L}_S h^{(0)}_A h^{(-3)}_B  \right).
    \end{split}
\end{align}
Obviously, by the same reasoning we can also introduce the difference of angular momenta between two cones. It can be obtain by changing the limits of the first integral -- from the whole real line to the interval $[u_1, u_2]$.  \\
Let us notice that a little bit more strict asymptotics (at large $u$) is needed to assure that the angular momentum are finite that in the case of the energy. One needs to have electric part of the Weyl tensor vanishing faster than $u^{-1}$ to guarantee convergent result. We relegate more detailed discussion on this topic to the Sec. \ref{allowed}.
\subsection{Momentum}\noindent
Derivation of momentum could follow the same line as the energy and angular momentum before. This calculation is straightforward but rather lengthy and unforgiving. For this reason we only give the final answer here for the total radiated momentum
\begin{equation}
   Q_{T_{(i)}} [h] =  \frac{1}{16\pi H } \int_{\mathcal{I}^+} d^3 x \sqrt{\mathring{q}}  \mathcal{E}_{cd} \left( \mathcal{L}_{T_{(i)}} h_{ab}^{(0)} - 2e^{Hu} H g_{i} h_{ab}^{(0)} \right) \mathring{q}^{ac} \mathring{q}^{bd} \label{momentum}
\end{equation}
and relegate all the details to the Appendix \ref{app_mom}. In the equation above, $\mathcal{E}$ is the electric part of the Weyl tensor whose explicit form is given in \eqref{weyl} and $\mathcal{L}_{T_{(i)}} h_{ab}^{(0)}$ is calculated in \eqref{translacje_pochodna_Lie}. 
As before, by changing limits of integration upon $u$ we can obtain the difference of momenta carried by two cones.
\section{Initial data} \label{allowed} \noindent
In this Section we will rewrite our charges in terms of the initial data $h_{ab}^{(0)}, \mathcal{E}_{ab}$, investigate what kind of boundary conditions (in the $u$ variable) one needs to assume to ensure that all the charges are finite. Furthermore, we will use those new expressions to compare with the charges previously calculated in \cite{Ashtekar:2015lla} and also discuss their gauge dependence.
\subsection{Boundary conditions} \noindent
Of course, all charges derived so far could be infinite if certain decay conditions as $u \to \pm \infty$ are not imposed. It is somehow a subtle issue. On one hand, one would like to have finite values of the physical charges. On the other hand, too restrictive boundary conditions can rule out physically interesting solutions. One would like at least to cover the same class as \cite{Ashtekar:2015lla}. We start with initial data considered there. Let $h_{ij}^P$ and $\mathcal{E}_{ij}^P$ denote perturbation of the metric and the electric part of the Weyl tensor\footnote{of course, we mean here and hereafter {\it linearized} electric part of the Weyl tensor.} induced on scri in the Poincar\'e patch respectively. Those are (up to an analogue of momentum constraints) free data \cite{friedrich1986}. \\  
We introduce spherical coordinates $(\mathcal{R}, x^A)$ in an usual manner. Background metric $\mathring{q}_{ij}^P$ reads:
\begin{equation}
    \mathring{q}^P = d\mathcal{R}^2 + \mathcal{R}^2 \mathring{\gamma}_{AB} dx^A dx^B
\end{equation}
Near the origin $\mathcal{R} = 0$, smooth perturbations behaves like
\begin{align}
\begin{split}
    h_{\mathcal{R} \mathcal{R}}^P &= O(1) \\
    h_{\mathcal{R} A}^P &= O(\mathcal{R}) \\
    h_{AB}^P &= O(\mathcal{R}^2) \label{smoothness}
    \end{split}
\end{align}
One can easily see that to go from $\mathbb{R}^3$ to $\mathbb{R} \times \mathbb{S}^2$, we need to change coordinates $\mathcal{R} = H^{-1} e^{-Hu}$. Then, background metrics are conformally equivalent:
\begin{equation}
    \mathring{q}^P = H^2\mathcal{R}^2 \left(du^2 + H^{-2} \mathring{\gamma}_{AB} dx^A dx^B \right) = H^2\mathcal{R}^2 \mathring{q}
\end{equation}
Obviously, perturbations must transform in the same manner:
\begin{align}
    \begin{split}
        h_{uu}^{(0)} &= h_{\mathcal{R} \mathcal{R}}^P \\
        h_{uA}^{(0)} &= -H^{-1} \mathcal{R}^{-1} h_{\mathcal{R} A}^P \\
        h_{AB}^{(0)} &= H^{-2} \mathcal{R}^{-2} h_{AB}^P.
    \end{split}
\end{align}
Taking into account \eqref{smoothness} we see hat $h^{(0)}$ goes to the constant as $u \to \infty$ (which corresponds to $\mathcal{R} \to 0$). It would follow that sub-leading terms would vanish exponentially. However, we simply demand that they behave as $u^{-\epsilon}$ for some $\epsilon > 0$.
In our gauge, only $h_{Au}^{(0)}$ is non-zero. $h_{AB}^{(-1)}$ and $h_{uu}^{(-1)}$ are given purely in terms of $\mathring{D}_B h_{Au}^{(0)}$ and thus satisfy the same asymptotics as $u \to \infty$. Let us mention for the comparison with the asymptotically flat case that $h_{AB}^{(-1)}$ of Christodoulou--Kleinerman solutions satisfies this decay condition with $\epsilon = \frac{1}{2}$ \cite{christodoulou1993global}. \\
The rest of the initial data are encoded in the electric part of the Weyl tensor $\mathcal{E}$ \cite{friedrich1986}. Under conformal transformations $\mathring{q} \mapsto \Omega^2 \mathring{q}$ it transform 'in the opposite way':
\begin{equation}
    \mathcal{E} \mapsto \Omega^{-1} \mathcal{E},
\end{equation} so we have:
\begin{align}
    \begin{split}
        \mathcal{E}_{uu} &= H^3 \mathcal{R}^3 \mathcal{E}_{\mathcal{R} \mathcal{R}}^P \\
        \mathcal{E}_{uA} &= - H^2 \mathcal{R}^2 \mathcal{E}_{\mathcal{R} A}^P \\
        \mathcal{E}_{AB} &= - H \mathcal{R} \mathcal{E}_{AB}^P.
    \end{split}
\end{align}
$\mathcal{E}^P$ should follow \eqref{smoothness} (if they are smooth), so we see that $\mathcal{E}$ vanishes as $e^{-3Hu}$. Let us express it through our perturbations:
\begin{align}
\begin{split}
    \mathcal{E}_{uu} &=  -H^3 h_{uu}^{(-3)} \\
    \mathcal{E}_{uA} &=   -\frac{3}{2}H^3 h_{A}^{(-3)} + H h_{A,u}^{(-2)} \\
    \mathcal{E}_{AB} &= - \frac{3}{2} H^3 h_{AB}^{(-3)} + \frac{1}{2}H h_{uu}^{(-3)} \mathring{\gamma}_{AB} - \frac{1}{H} \mathring{D}_{(B} h_{A)}^{(0)} - H \mathring{D}_{(B} h_{A)}^{(-2)} - \frac{1}{2H}\mathring{D}_A \mathring{D}_B h_{uu}^{(-1)} - \frac{1}{2H}h_{AB,uu}^{(-1)}. \label{weyl}
\end{split}
\end{align}
A short calculation shows that $\mathring{q}^{ab} \mathcal{E}_{ab} = 0$ as one would expect.
For the case of Schwarzschild--de Sitter solution, one easily obtains only $\mathcal{E}_{uu} = \textrm{const.}  \neq 0$. \\
Let us now rewrite our fluxes using $\mathcal{E}$. After an integration by parts on a sphere (so no boundary terms occur), we obtain:
\begin{align}
\begin{split}
    Q_T &= \frac{1}{16\pi H}\int_{\mathcal{I}^+} d^3x \sqrt{\mathring{q}} \mathcal{E}_{cd} \mathcal{L}_T \left(H^{-2}h^{(0)}_{ab}\right) \mathring{q}^{ac} \mathring{q}^{bd} \\
    Q_S &=  \frac{1}{16 \pi H} \int_{\mathcal{I}^+} d^3 x \sqrt{\mathring{q}} \mathcal{E}_{cd} \mathcal{L}_{S} \left(H^{-2}h_{ab}^{(0)} \right) \mathring{q}^{ac} \mathring{q}^{bd}
\end{split} \label{ladunki_elekt}
\end{align}
and we see that much less restrictive conditions are needed to be imposed upon $\mathcal{E}$ to ensure their finite values as $u \to \infty$. Indeed, one can easily notice that the energy density is integrable as $u \to \infty$ if simply $\mathcal{E}_{uA}$ has a finite limit. To ensure convergence of $Q_S$ we need to assume a little bit more, namely that it goes to zero faster than $u^{-1}$. Finally, to ensure finite momenta, we need to impose condition that $e^{Hu} \mathcal{E}$ is integrable at infinity as seen from \eqref{momentum}. Since $\mathcal{L}_{T_{(i)}} h^{(0)}$ has not only $uA$ components, it is restriction on the whole $\mathcal{E}$. Remarkably, it is still milder condition then the one implied by \cite{Ashtekar:2015lla}. \\
An attentive reader may be worried about boundary terms which were abandoned when we transited from the definition of the energy emitted in the interval $[u_1, u_2]$ and the total energy. Indeed, if we assume simply that $\mathcal{E}$ has a finite limit, then they give a contribution to $Q_T$. Since one would like to have also $Q_S$ and $Q_{T_{(i)}}$ well-defined, it is not problematic and they indeed vanish. \\
Finally, let us explain perhaps a cumbersome way \eqref{ladunki_elekt} is written, one should notice that induced perturbation of the metric on $\mathcal{I}^+$ is not $h_A^{(0)}$ but rather $H^{-2} h_A^{(0)}$.
\\ 
An attentive reader definitely noticed that we were so far concerned only with the $u \to \infty$ limit, neglecting $u \to -\infty$. In this setting they are physically very distinct. The former corresponds to the $i^+$ where the observer hits the {\it scri} whereas the former corresponds to the observer's cosmological horizon. Thus one could impose different conditions at both ends. However, from the point of view of both $Q_T$ and $Q_S$ it is not needed because those expressions are symmetric with respect to the $u \mapsto -u$. $Q_{T_{(i)}}$ breaks this symmetry but due to $e^{Hu}$ factor in the definition of $T_{(i)}$ \eqref{translacje}, integrand vanishes quickly as $u \to - \infty$ and is not problematic. Thus, we can impose the same decay at both ends. In the language of the Poincar\'e data, it means that they obey power law behavior at large $\mathcal{R}$.
\subsection{Comparison} \noindent
We will now compare our results with those obtained in \cite{Ashtekar:2015lla}. It is easiest done using the form \eqref{ladunki_elekt} because we already know how to transform both $h^P$ and $\mathcal{E}^P$. Let us remind the form of charges obtained therein
\begin{align}
    \begin{split}
            Q_T^P &= \frac{1}{2H\kappa}\int_{\mathcal{I}^+} d^3x \sqrt{\mathring{q}^P} \mathcal{E}_{ij}^P \left(\mathcal{L}_T h^P_{kl} + 2H h^P_{kl} \right) \mathring{q}^{ikP} \mathring{q}^{jlP} \\
            Q_{T_{(i)}}^P &=  \frac{1}{2 H \kappa} \int_{\mathcal{I}^+} d^3 x \sqrt{\mathring{q}^P} \mathcal{E}^P_{ij} \mathcal{L}_{T_{(i)}} h_{kl}^P \mathring{q}^{ikP} \mathring{q}^{jlP} \\
            Q_{S}^P &=  \frac{1}{2 H \kappa} \int_{\mathcal{I}^+} d^3 x \sqrt{\mathring{q}^P} \mathcal{E}_{ij}^P \mathcal{L}_{S} h_{kl}^P \mathring{q}^{ikP} \mathring{q}^{jlP}. \label{poincare_charges}
    \end{split}
\end{align}
(One should keep in mind that there is also a factor of $\det \mathring{q}^P$ hidden in the measure in formulas given in \cite{Ashtekar:2015lla}.) Let us now express it using $h^{(0)}$ and $\mathcal{E}$ by the means of the conformal transformation discussed above. We immediately find: 
\begin{align}
\begin{split}
    Q_T^P &= \frac{1}{2H\kappa}\int_{\mathcal{I}^+} d^3x \sqrt{\mathring{q}} \mathcal{E}_{cd} \mathcal{L}_T\left( H^{-2} h^{(0)}_{ab} \right) \mathring{q}^{ac} \mathring{q}^{bd} \\
        Q_{T_{(i)}}^P &=  \frac{1}{2 H \kappa} \int_{\mathcal{I}^+} d^3 x \sqrt{\mathring{q}}  \mathcal{E}_{cd} H^{-2}\left( \mathcal{L}_{T_{(i)}} h_{ab}^{(0)} - 2e^{Hu} H g_{i} h_{ab}^{(0)} \right) \mathring{q}^{ac} \mathring{q}^{bd} \\
    Q_{S}^P &=  \frac{1}{2 H \kappa} \int_{\mathcal{I}^+} d^3 x \sqrt{\mathring{q}} \mathcal{E}_{cd} \mathcal{L}_{S} \left( H^{-2} h_{ab}^{(0)} \right) \mathring{q}^{ac} \mathring{q}^{bd}
\end{split} \label{porownanie}
\end{align}
which are exactly equal to \eqref{ladunki_elekt} if one remembers that $\kappa = 8\pi$ in our units. 
Of course, it is not a surprise. We used exactly the same definitions for fluxes after all. However, as discussed above, our results apply to a broader class of solutions.
\subsection{Gauge invariance} \noindent
The form \eqref{ladunki_elekt} is extremely convenient to discuss gauge invariance of our results. For concreteness we will focus on $Q_T$. Gauge transformations preserving Bondi gauge can be realized as 
\begin{align}
\begin{split}
    h_{ab}^{(0)} &\mapsto  h_{ab}^{(0)} + H^2 \mathring{\nabla}_{(a} \xi_{b)} \\
    \mathcal{E}_{ab} &\mapsto \mathcal{E}_{ab},
    \end{split}
\end{align}
where $\xi$ is tangent to $\mathcal{I}$. Moreover, it satisfies\footnote{Interestingly, not all residual gauge transformations \eqref{res_gen} are of this form but this will not affect our reasoning.} 
\begin{align}
    \begin{split}
        \mathring{\nabla}_u \xi_u &= 0 \\
        \mathring{D}_{(A} \xi_{B)} & = 0. \label{trans}
    \end{split}
\end{align}
Derivation of residual gauge freedom is presented in the App. \ref{residual} -- their final forms are presented as \eqref{res_gen} and \eqref{final res}. \\
Let us now calculate how energy radiated between $u_1$ and $u_2$ transforms:
\begin{equation}
    E(u_1) - E(u_2) \mapsto E(u_1) - E(u_2) + \frac{1}{16\pi H^3}\int_{u_1}^{u_2} \int_{\mathbb{S}^2} dx^A \sqrt{\mathring{\gamma}} \mathcal{E}_{cd} \mathcal{L}_T \mathring{\nabla}_a \xi_b \mathring{q}^{ac} \mathring{q}^{bd}
\end{equation}
Since $T$ is a Killing vector of $\mathring{q}$, we have
\begin{equation}
    [\mathcal{L}_T, \mathring{\nabla}_a] = 0
\end{equation}
Moreover, $\mathcal{E}$ is divergence-free and thus
\begin{equation}
   \frac{1}{16\pi H^3}\int_{u_1}^{u_2} \int_{\mathbb{S}^2} dx^A \sqrt{\mathring{\gamma}} \mathcal{E}_{cd} \mathcal{L}_T \mathring{\nabla}_a \xi_b \mathring{q}^{ac} \mathring{q}^{bd} = \frac{1}{16\pi H^3}\int_{u_1}^{u_2} \int_{\mathbb{S}^2} dx^A \mathring{\nabla}_a \left( \sqrt{\mathring{\gamma}} \mathcal{E}_{cd} \mathcal{L}_T \xi_b \mathring{q}^{ac} \mathring{q}^{bd}  \right)
\end{equation}
and so it produces boundary terms:
\begin{equation}
    \frac{1}{16\pi H^3} \int_{\mathbb{S}^2} dx^A \sqrt{\mathring{\gamma}} \mathcal{E}_{ad} \mathcal{L}_T \xi^d   (\partial_u)^a |_{u_1}^{u_2} = \frac{1}{16\pi H^3} \int_{\mathbb{S}^2} dx^A \sqrt{\mathring{\gamma}} \mathcal{E}_{ud} \xi^d_{\ ,u} |_{u_1}^{u_2} \label{energy_transformed}
\end{equation}
Since we assume that $\mathcal{E}$ vanishes as $u \to \pm \infty$, $Q_T$ as a whole is gauge invariant. Unfortunately, distribution between $u_1$ and $u_2$ is not. This term vanishes\footnote{It is somehow non-obvious how one should calculate orders of magnitude. If we want to impose that $h^{(0)}$ is proportional to $\Lambda$ even after transformation, then $\xi_a$ is $O(1)$ and $\xi^a$ not.} as $\Lambda \to 0$. However, if one classifies all possible divergences which are trilinear in $h^{(0)}$, $\mathcal{E}$ and $T$, one immediately finds that we cannot add anything to our density to make it gauge invariant. However, situation changes drastically if we further assume the rigid transport condition \eqref{rigid_transport}, we immediately finds that \eqref{energy_transformed} vanishes. Indeed, it works on the initial data space through $\xi = f(x^A) \partial_u$ and so 
\begin{equation}
    \mathcal{L}_T \xi = 0.
\end{equation}
which makes quasilocal difference of energies invariant. As follows from our analysis, it is the best one can get -- it is impossible to obtain expression invariant under all residual transformations \eqref{res_gen}. Thus we propose the following definition of the emitted energy in the interval $[u_1, u_2]$:
\begin{align}
&E(u_1)-E(u_2)=\nonumber\\
&\frac{1}{32 \pi}\int_{u_1}^{u_2} du\int_0^\pi \sin \theta d \theta\int_0^{2\pi} d\phi \mathring{\gamma}^{AB} \left( \mathring{\gamma}^{CD} h^{(-1)}_{AC,u} h^{(-1)}_{BD,u} - 6 h^{(-3)}_{A} h^{(0)}_{B,u} 
    \right)
\end{align}
as the only expression with the correct $\Lambda \to 0$ limit and invariant under superpseudotranslations.
One recognizes that it is exactly \eqref{chrusciel}.
\section{de Sitter energy is not translationally invariant} \label{komutatory}
\noindent
In this section we will first discuss the general transformation properties of charges defined on a phase space. As a particular example, we will take de Sitter group. Later on, we will check whether and how those properties are implemented in topologically different scenarios (to be more precise -- how it depends upon the topology of the {\it scri} $\mathcal{I}^+$) distinguished in \cite{Ashtekar:2014zfa}. We will follow terminology introduced therein.
\subsection{General setting} \noindent
Let us assume that our theory is equipped with symmetries generated by an algebra $\mathcal{A}$. Furthermore, we assume they are represented on the phase space $\left(\mathcal{P}, \lbrace \cdot, \cdot \rbrace\right)$ by charges $Q_S$:
\begin{equation}
    \lbrace Q_S, \cdot \rbrace = -\delta_S
\end{equation}
for any $S \in \mathcal{A}$. Let $O \in C^{\infty}(\mathcal{P})$ be any function defined on the phase space. Let $S, T \in \mathcal{A}$. From the Jacobi identity we have
\begin{equation}
    -\lbrace \delta_S Q_T, O \rbrace =  \lbrace \lbrace Q_S, Q_T \rbrace, O \rbrace = \lbrace Q_S, \lbrace Q_T, O \rbrace \rbrace + \lbrace Q_T, \lbrace O, Q_S \rbrace \rbrace
\end{equation}
and from antisymmetry
\begin{equation}
   \lbrace \delta_S Q_T, O \rbrace = \delta_S \delta_T O - \delta_T \delta_S O = \delta_{[S,T]} O = -\lbrace Q_{[S,T]}, O \rbrace.
\end{equation}
Since $O$ is arbitrary, we see that
\begin{equation}
    \delta_S Q_T = -Q_{[S,T]} + \textrm{const.}
\end{equation}
If there is a maximally-invariant vacuum (as is the case in our considerations), we can put this constant to zero automatically.
We conclude that $\delta_S Q_T \neq 0$ as long as $[S,T] \neq 0$. \\
In particular, one could reasonably assume that any theory describing asymptotically de Sitter spacetimes (in particular, any linear theory on the background of the de Sitter) should have de Sitter group among its (possibly larger) symmetries. \\
This group contains a three dimensional abelian subgroup which can be identified with translations (in particular they are translations in the Poincar\'e patch \cite{Ashtekar:2015lla}). However, their generators $T_{(i)}$ do not commute with a 'time-translation' generator $T$:
\begin{equation}
    [T, T_{(i)}] = H T_{(i)},
\end{equation}
where $H$ is the Hubble constant. Thus,
\begin{equation}
    \delta_{T_{(i)}} Q_T = H Q_{T_{(i)}} \label{wariacja}
\end{equation}
$Q_T$ was identified (e.g. in \cite{Ashtekar:2015lla}) with the energy carried by GWs which leads to the conclusion that it is not a translationally-invariant concept, at least for generic perturbations.  
\subsection{Globally asymptotically de Sitter} \noindent
It is well-known fact that any linear perturbation on the {\it whole} de Sitter space must have vanishing charges of the de Sitter group because Cauchy surfaces are compact  \cite{Moncrief:1976un}. Thus, GWs carry no total energy nor momentum and \eqref{wariacja} is trivially satisfied. However, usually we have only an access to the part of the spacetime (bounded by our cosmological horizon). Charges of such restricted data are no longer constrained by this requirement.
\subsection{Asymptotically de Sitter in a Poincar\'e patch} \noindent
It was proposed in \cite{Ashtekar:2015lla} how to calculate charges associated with 7 de Sitter isometries preserving given Poincar\'e patch (it means, tangent to the previously chosen cosmological horizons). They were identified as momenta, angular momenta and energy. The last one is probably the most controversial due to the fact that the associated Killing vector $T$ is spacelike near {\it scri}. Those charges are given by formulas \eqref{poincare_charges}. From these expressions it is an easy calculation to see that\footnote{Following \cite{Ashtekar:2015lla}, we assumed that both $h$ and $\mathcal{E}$ are of Schwartz class at $\mathcal{I}^+$ so we could neglect boundary terms while integrating by parts.}
\begin{align}
\begin{split}
    \delta_{T_{(i)}} Q_T^P &= \frac{1}{2H \kappa} \int_{\mathcal{I}^+} d^3 x \sqrt{\mathring{q}^P} \mathcal{E}_{ij}^P \mathcal{L}_{[T,T_{(i)}]} h_{kl}^P \mathring{q}^{ikP} \mathring{q}^{jlP} \\ &= \frac{1}{2 \kappa} \int_{\mathcal{I}^+} d^3 x \sqrt{\mathring{q}^P} \mathcal{E}_{ij}^P \mathcal{L}_{T_{(i)}} h_{kl}^P \mathring{q}^{ikP} \mathring{q}^{jlP} = H Q_{T_{(i)}}^P.
    \end{split}
\end{align}
The geometrical origin of this law is quite simple. Vector field $T$ is not fixed unambiguously at $\mathcal{I}^+$. It is distinguished by the additional condition that it leaves $i^+$ invariant. However, since $T_{(i)}$s do not  have this property, upon their action there is different $i^+$ and thus also different $T$. \\
This result raises an important question: energy is supposed to be very physical concept, it could be used to warm up a cup of tea. Is this lack of the translational invariance a sign that in fact expression \eqref{poincare_charges} cannot be interpreted as energy one could use? To answer this question we will consider two different scenarios. 
\subsubsection{Physical interpretation} \noindent
The simpler one concerns itself with globally-defined solutions to the sourceless linearized Einstein equations. Such solutions were accused of being non-physical \cite{Ashtekar:2015lla} because their $Q_T$ is unbounded from below but nevertheless they provide a~useful example. Picturesquely, we can imagine that we have a wave moving e.g. to the right (it means it has a non-vanishing momentum in this direction) and we move it, also to the right by some distance. However, since our wave moves to the right, this procedure could be (roughly speaking) seen as translation not in spatial direction but in the conformal time $\eta$. This is even more clear when one consider how $T_{(i)}$ acts on the cosmological horizon. 
But our background describes an expanding universe. It is not in the least surprising that waves crossing our horizons later carry less energy than similar ones entering earlier on. Incidentally, let us notice that the conformal energy introduced in \cite{kastor2002positive} is translationally invariant for the simple reason that
\begin{equation}
    [\partial_\eta, \partial_{x^i}] = 0,
\end{equation}
where $(\eta, x^i)$ are the usual coordinates used in the Poincar\'e patch. \\
The second scenario (physical one) is concerned with GWs produced in some astrophysical event, e.g. in the merger of a binary system.  This time, our translation moves not only wave but also source and we cannot simply translate it into a  conformal time translation. However, let us notice that our analysis is not restricted to the GWs but all carriers of de Sitter charges, including all final products of the merger. If the initial (spacelike) momentum was zero, $Q_T$ of the whole system is the same, the only thing that differs is the way it distributed between different products (for concreteness, let us assume it were GWs and a~black hole). It is by no means a new effect. Indeed, it was already noticed in 1965 that the split into Coulombic and radiative modes is origin dependent \cite{penrose1965zero}. Thus, we are led to the believe that energy released in such a process can have well-defined, physical meaning as long as one does not try to associate it only with radiation.  \\
In the \cite{Ashtekar:2015lxa} the quadrupole formula for such situations was derived and found to be positive. However, our analysis shows that it could be arbitrarily negative if one moved the source because one would use $T$ not adapted to it anymore. Let us notice that any (real-life) detection of GWs carries much more information than a few values of (Minkowski) charges. It seems conceivable that having a full wavefront, one could extract some 'translationally-invariant' part of the energy (it means, combined energy of the GW and BH) which (hopefully) were positive. This issue gets even more troublesome when we have more than two sources and thus there is no distinguished $T$. We plan to investigate it in the future.
\subsection{Asymptotically Schwarzschild--de Sitter}
\noindent
Finally, we can analyse the case of the asymptotically Schwarzschild--de Sitter spacetime in which {\it scri} is topologically a cylinder $\mathbb{S}^2 \times \mathbb{R}$, it means $\mathbb{R}^3$ with one point removed. This point ($i^+$) corresponds to the intersection of {\it scri} with a fixed line (either an privileged observer or a black hole). Since translations changes such a line, they also move 'a point at infinity'. Geometrically speaking, $T_{(i)}$ are tangent to $\mathcal{I}^+$ but are not complete vector fields. It is immediate to see that this case can be obtained working in the Bondi coordinates \eqref{bondi_exp} and using $\Omega = H^{-1} r^{-1}$ as a conformal factor. \\ 
It is not obvious how translations should act upon phase space in this case. As we mentioned, they are not complete and thus can ruin any reasonable boundary conditions (e.g. assumption that the initial data have power-law decay as is clear from the \eqref{translacje}). For simplicity, one could restrict themselves to the compactly supported initial data because this property is preserved by the {\it infinitesimal} translations. This class was actually already covered in the previous subsection as follows from our discussion about relationship between Poincare patch and Bondi coordinates. Moreover, we started with \eqref{wariacja} to derive the form of $Q_{T_{(i)}}$ and so it is tautologically satisfied in this paper.
\section{Energy flux through the Killing horizon} \label{horizon} 
\noindent
It was recently suggested that it is more natural to describe gravitational radiation in the region of spacetime bounded by two intersecting null surfaces rather than {\it scri} \cite{Ashtekar:2019khv}. This idea seems to be geometrically appealing because such system looks just like asymptotically flat spacetime on the Penrose diagram. In the case of the de Sitter background (in the Poincar\'e patch) those two surfaces can be naturally chosen as a bifurcated Killing horizon (of which, one part is a~cosmological horizon)\footnote{Obviously, there are many Killing horizons in the de Sitter universe. Thus, one needs to choose Poincar\'e patch first to distinguish it}. 
Then, it can be argued that in the limit $\Lambda \to 0$ those surfaces become the future and past, respectively,  {\it scri} of the final Minkowski spacetimes of the final Minkowski spacetimes. Indeed, they already are equipped with the correct signature and topology.
Although this line of reasoning is very convincing, it restricts itself to the geometry of background and not concerns dynamical fields living upon it. The energy flux through the horizon was previously calculated in \cite{Date:2016uzr, Hoque:2018byx}. However, authors used different gauge and did not consider the limiting procedure so we will repeat the whole process here. In this Section we  discuss how the limit $\Lambda \to 0$ looks like from the point of linearized gravity nearby this Killing horizon. \\
We focus on the future horizon
$$ {\cal H}:= \Sigma_{r_c}, \ \ \ \ \ \ \  r=\sqrt{\frac{3}{\Lambda}}$$
which replaces now the {\it scri} $\mathcal{I}^+$. Physically relevant part of that horizon lies between the bifurcation sphere and the {\it scri}. In the terms of our Bondi frame,  the physically relevant part of spacetime is
$$ r\le r_c. $$
Therefore, we consider now the suitable segments of the cones null cones $\tilde{C}_{u_1}$ and $\tilde{C}_{u_2}$, and  the connecting segment o
$\Sigma_{r_c}(u_2,u_1)$, of the horizon. The radiated energy balance becomes
\begin{equation}
\tilde{E}(u_1)-\tilde{E}(u_2) = -\frac{1}{2}\int_{\Sigma_{r_c}(u_2,u_1)}\omega^a(h,{\cal L}_Th)\epsilon_{abcd}\frac{1}{3!}dx^bdx^cdx^d \label{roznica_hor}
\end{equation}
One can calculate the flux through the horizon using the right hand side of \eqref{roznica} evaluated at $r=r_c$ instead of infinity. The symplectic current at the horizon yields: 
\begin{align}
    &-16 \pi \omega^r (h_1, h_2) =\nonumber\\ 
&h_{2BC} \nabla_D h_{1uA} g^{CD} g^{AB} - h_{2Bu} \nabla_r h_{1uA} g^{AB} - \frac{1}{2} h_{2BD} \nabla_u h_{1AC} g^{AB} g^{CD} \nonumber\\&- 1 \longleftrightarrow 2  \label{horizon_flux}
\end{align}
The formula (\ref{omegar}) also applies,  what is special about the horizon case is
$$ g^{rr}_{|_{r=r_c}}=0 $$
hence the first row in (\ref{omegar}) disappears.  
The resulting  formula for the radiated energy flux through the horizon becomes particularly simple if the vectors $\partial_u$ and $\partial_A$ are still orthogonal with respect to the perturbed metric tensor $g+h$, namely, in that case 
$$h_{uA}=0$$
and
 \begin{align}     \tilde{E}(u_1)-\tilde{E}(u_2) = \frac{1}{32 \pi r^2_c}\int_{u_1}^{u_2} du\int_0^\pi \sin \theta d \theta\int_0^{2\pi} d\phi \nonumber\\
       \left(\mathring{\gamma}^{BE} \mathring{\gamma}^{FC} h_{BC,u}h_{EF,u}  - \frac{1}{2}\partial_u\left(\mathring{\gamma}^{BE}  \mathring{\gamma}^{FC} h_{BC}h_{EF,u}\right)\right)
\end{align}
Remarkably, we could make that assumption without loss of generality, as a~different gauge fixing condition instead of (\ref{bondi_exp}) and (\ref{gauge}) while assuming the second condition in (\ref{gauge}) only at the horizon.  Then $h_{AB,u}$ is free data defined at the horizon.
We will now consider the $\Lambda \to 0$ limit of \eqref{energy_flux} but to this end we need to go back to the gauge conditions (\ref{gauge}) and the expansion (\ref{bondi_exp}) and assume a little bit more about our solutions. Indeed, let us assume that the Bondi expansion \eqref{bondi_exp} of our solution is convergent near $r = r_c$ in a variable $r^{-1}$. Moreover, we assume that this still holds when we vary $\Lambda$ (while taking initial data in the way explained in the Footnote 3). Since our theory is linear, every coefficient is going to depend upon $h_{AB}^{(-1)}$ (and some other freely specified data like $h^{(-3)}_A$) linearly. Moreover, those linear relationships are holomorphic in $\Lambda$ around $\Lambda = 0$\footnote{It could, and in fact does happen that $\Lambda$ multiplies a new term in the expansion (for an example, see Eq. (2.20) in \cite{Compere:2019bua}. However, in all such situations it can only impose a homogeneous constraint. Otherwise, Einstein equations linearized around Minkowski were singular.}. Thus, all $h_{AB}^{(-n)}$ and $h_{B}^{(-n)}$ are polynomials in $\Lambda$. Since $r_c = \sqrt{\frac{3}{\Lambda}}$ and the Bondi expansion is a~Laurent series in $r^{-1}$ perturbations components evaluated at the cosmological horizon are also Laurent series in $\sqrt{\Lambda}$. Assuming this series to be convergent for $\Lambda > 0$, we can use it to calculate the limit $\Lambda \to 0$ of \eqref{horizon_flux} obtaining:
\begin{equation}
    \frac{1}{32 \pi}\int_{\mathcal{I}^+} du \sin \theta d \theta d\phi \mathring{\gamma}^{AB} \mathring{\gamma}^{CD} \stackrel{(-1)}{h}_{AC,u} \stackrel{(-1)}{h}_{BD,u}
\end{equation}
which is exactly Trautman--Bondi formula. Details are presented in the Appendix \ref{app_hor}. \\
A few comments are in place:
\begin{itemize}
    \item[(i)] It may be seen as a trivial exercise since the Minkowski limit was already established in the Sec. \ref{bondi}. However, this is not the case because there is yet another potential source of the energy through {\it scri}, namely (past) cosmological horizon above the bifurcation sphere. This results shows that if one takes the limit $\Lambda \to 0$ in an appropriate manner, those contributions are getting smaller and smaller which somehow strengthens the idea that we can physically restrict our considerations to the region bounded by two null surfaces. As a byproduct, it seems that our method of taking the limit is consistent with the 'no-incoming-radiation' proposal of \cite{Ashtekar:2019khv}.
    \item[(ii)] Although this result supports those ideas and proposals, our line of reasoning in fact uses additional structures (like the behaviour of solutions near $\mathcal{I}^+)$. Unfortunately, we do not know how to take this limit intrinsically at the horizon. \\
    It seems to us that the gauge choices of \cite{Hollands:2012sf} are better suited for the task (and indeed, something that {\it looks like} the Trautman--Bondi formula in the Bondi gauge follows immediately even at finite $\Lambda$ despite the fact it is null Gau\ss ian gauge) but it is less clear how data defined on the horizon translates into those on the Minkowski {\it scri}. \\
    As noted in \cite{Compere:2019bua}, not all vacuum solutions can be obtained in the limit $\Lambda \to 0$, at least those which exhibit terms logarithms in $r$ are explicitly excluded. Although their physical status is not clear (since they do not enjoy a smooth conformal completion), it would be of interests to understand if they can be recovered by admitting a broader class of initial data at the horizon.
    \item[(iii)] This derivation relies heavily upon the analytical properties of solutions both in $r^{-1}$ and in $\sqrt{\Lambda}$. We hope to investigate if they are truly satisfied in the future. Let us notice that as $\Lambda \to 0$, needed radius of convergence in $r^{-1}$ is getting smaller so those assumptions seem to be reasonable.
    \item[(iv)] It may be useful for future uses to note how data induced on the horizon behaves for small $\Lambda$. We have:
    \begin{align}
    \begin{split}
        h_{AB} &= h_{AB}^{(-1)} \sqrt{\frac{3}{\Lambda}} + O(\Lambda^{\frac{1}{2}}) \\
        h_B &= -P^{AC}_{\ \ B} h_{AC}^{(0)} + \frac{1}{2} \mathring{D}^A h_{AB}^{(-1)} + O\left(\Lambda^{\frac{1}{2}} \right), 
        \end{split} \label{granice}
    \end{align}
    where $P^{AC}_{\ \ \ B}$ is an (non-unique) inverse operator of the one at the left hand side of \eqref{constraint} and we assumed that the electric part of the Weyl tensor is of order $O(1)$. \\
    One should also notice that this expansion is not equivalent to the simple $r_c \to 0$ limit but also scaling of different coefficients with $\Lambda$ matters. Otherwise, terms with $h^{(0)}_B$ would survive. 
\end{itemize}
\section{Conclusions} \label{conclusions} \noindent
{In this work we have investigated a few aspects of GWs in the de Sitter background. The key factor which united them all was the usage of the symplectic approach. \\
First of all, following \cite{Chrusciel:2020rlz} we have  derived the de Sitter spacetime generalization of the Trautman-Bondi energy loss formula and also calculated fluxes of angular momentum and momentum. In particular, we have derived de Sitter charges for linearised gravity at $\mathcal{I}^+$ in the Bondi gauge and expressed them in terms of the initial data $h_{ab}^{(0)}$ and $\mathcal{E}_{ab}$. We compared them with those obtained in \cite{Ashtekar:2015lla} finding that expressions are equivalent but our gauge allows for a broader class of solutions. We discussed gauge transformations finding that our expressions for the quasilocal energy are not gauge invariant but the total energy is. It remains an open problem whether one can modify this recipe to obtain invariant quantity which at the same time possess the correct limit $\Lambda \to 0$.
In the Sec. \ref{komutatory} we discussed how our charges should transform under the action of the de Sitter group finding non-trivial behavior of the energy under translations generated by $T_{(i)}$. We tried to clarify this issue on a few examples. It seems to us that this topic is of relevance especially for the processes in which there is more than one compact object at the end and then there is no distinguished $T$ and is thus tightly connected to the theory of $\mathcal{S}$-matrix in the de Sitter universe. Finally, in the Sec. \ref{horizon}  we derived the energy flux associated with the Killing vector $T$ for the part of spacetime bounded by the future cosmological horizon. Having assumed that the Bondi expansion \eqref{bondi_exp} is still valid, we were able to find $\Lambda \to 0$ limit of the fields in question and of the flux itself. Remarkably, it reproduces Trautman--Bondi mass loss formula and thus strengthens the idea that null surfaces are more suited to serve as boundaries for the astrophysical processes recently proposed in \cite{Ashtekar:2019khv}.
\\
It seems to us that although there is still a lot to understand about GWs in the expanding universe, we are on the right track in the search for the appropriate framework to discuss radiative phenomena. 
}
\begin{acknowledgements} \noindent
We thank Piotr Chruściel, Sk Jahanur Hoque and Tomasz Smołka for useful informations regarding their paper and Abhay Ashtekar for fruitful discussions. MK was financed from budgetary funds for science in 2018-2022 as a research project under the program "Diamentowy Grant". JL was supported by Project OPUS 2017/27/B/ST2/02806 of  Polish National Science Centre.
\end{acknowledgements}
\appendix
\section{Energy -- derivation} \label{app}
\subsection{\it{Scri}} \label{app_scri} \noindent
Evaluation of \eqref{sympl} in our gauge gives
\begin{align}
    \begin{split}
        16 \pi\omega^r(h, \mathcal{L}_{\partial_u} h) = &- \frac{1}{2}g^{rr} g^{BE} g^{FC} \left(\mathcal{L}_{\partial_u}h_{BC} \partial_r h_{EF} - h_{BC} \partial_r \mathcal{L}_{\partial_u} h_{EF}  \right) \\
        &+g^{FB} \left(\mathcal{L}_{\partial_u}h_{Bu} \partial_r h_{uF} - h_{Bu} \partial_r \mathcal{L}_{\partial_u} h_{uF} \right) \\
         &-g^{FB}g^{CD} \left(\mathcal{L}_{\partial_u}h_{BC} \partial_D h_{uF} - h_{BC} \partial_D \mathcal{L}_{\partial_u} h_{uF} \right) \\
        &+ \frac{1}{2} g^{BE}g^{FC} \left(\mathcal{L}_{\partial_u}h_{BC} \partial_u h_{EF} - h_{BC} \partial_u \mathcal{L}_{\partial_u} h_{EF} \right)\\
        &+g^{FB} g^{CD} \Gamma_{\ DF}^H \left(  \mathcal{L}_{\partial_u}h_{BC} h_{Hu} - h_{BC} \mathcal{L}_{\partial_u} h_{Hu} \right). \label{omegar}
    \end{split}
\end{align}
Let us now consider the case $r \to \infty$ and check asymptotics. Using \eqref{bondi_exp} and constraint \eqref{constraint} we can estimate all terms:
\begin{equation}
    - \frac{1}{2}g^{rr} g^{BE} g^{FC} \left(\mathcal{L}_{\partial_u}h_{BC} \partial_r h_{EF} - h_{BC} \partial_r \mathcal{L}_{\partial_u} h_{EF}  \right) = o(r^{-2}),
\end{equation}
\begin{equation}
    g^{FB} \left(\mathcal{L}_{\partial_u}h_{Bu} \partial_r h_{uF} - h_{Bu} \partial_r \mathcal{L}_{\partial_u} h_{uF} \right) =-\frac{3}{r^2} \mathring{\gamma}^{FB} \left(h^{(-3)}_{B,u} h^{(0)}_{F} - h^{(-3)}_{B} h^{(0)}_{F,u} \right) + o(r^{-2}),
\end{equation}
\begin{equation}
    -g^{FB}g^{CD} \left(\mathcal{L}_{\partial_u}h_{BC} \mathring{D}_D h_{uF} - h_{BC} \mathring{D}_D \mathcal{L}_{\partial_u} h_{uF} \right) = O(r^{-3}),
\end{equation}
\begin{equation}
    \frac{1}{2} g^{BE}g^{FC} \left(\mathcal{L}_{\partial_u}h_{BC} \partial_u h_{EF} - h_{BC} \partial_u \mathcal{L}_{\partial_u} h_{EF} \right) = \frac{1}{2r^2} \mathring{\gamma}^{BE} \mathring{\gamma}^{FC} \left(h^{(-1)}_{BC,u} h^{(-1)}_{EF,u} - h^{(-1)}_{BC} h^{(-1)}_{EF,uu} \right) + o(r^{-2})
\end{equation}
And so, after some integration by parts:
\begin{align}
    \begin{split}
        16\pi\omega^r &= \frac{1}{r^2} \mathring{\gamma}^{BE}  \left(\mathring{\gamma}^{FC} h_{BC,u}^{(-1)} h_{EF,u}^{(-1)} - 6 h^{(-3)}_B h^{(0)}_{E,u}\right)  \\
        &- \frac{1}{2r^2} \mathring{\gamma}^{BE} \partial_u \left(\mathring{\gamma}^{FC} h_{BC}^{(-1)} h_{EF,u}^{(-1)}  - 6h^{(-3)}_B h^{(0)}_E\right) + o(r^{-2}) \label{finite}
    \end{split}
\end{align}
Since the measure contains factor $r^2$ integral of $\omega^r$ is finite and we obtain \eqref{energy_flux}.
\subsection{Horizon} \label{app_hor} \noindent
Let us dwell into details of taking the limit $\Lambda \to 0$ at the horizon. First of all, since the Killing vector $\partial_u$ is $\Lambda$-independent, we can focus on the general form of the symplectic current, taking $h_2 = \mathcal{L}_{\partial_u} h_1$ at the very end. As we mentioned in the main text, we assume that the expansion \eqref{bondi_exp} is convergent nearby the horizon (so that we can differentiate under the sum) and moreover, that when evaluated at the horizon it is also a convergent power series in $\sqrt{\Lambda}$ (so that we can take the limit $\Lambda \to 0$ easily). We need to estimate $\omega^r$ (evaluated at $r=r_c$). Let us notice that it is integrated together with the factor $r_c^2 = \frac{3}{\Lambda}$ and thus all factors $o(\Lambda)$ do not contribute in the limit. Of course, one needs to check also that all factors of $O(1)$ cancel out to ensure finite result. \\
We have from \eqref{bondi_exp}:
\begin{equation}
    h_{AB}(r=r_c) = h^{(-1)}_{AB} \sqrt{\frac{3}{\Lambda}} + \frac{h^{(-3)}_{AB}}{\sqrt{\frac{3}{\Lambda}}} +...
\end{equation}
Since all higher order coefficients are at least of order $(1)$, we obtain 
\begin{equation}
    h_{AB}(r=r_c) = h^{(-1)}_{AB} \sqrt{\frac{3}{\Lambda}} + O(\sqrt{\Lambda}).
\end{equation}
Analogously:
\begin{equation}
    h_{A}(r=r_c) = h_{A}^{(0)} \frac{3}{\Lambda} + \frac{1}{2} \mathring{D}^B h_{AB}^{(-1)} +... 
\end{equation}
If we were simply working with the large $r$ limit, the first term would be dominant. However, one needs to remember that in fact $h_{A}^{(0)}$ is proportional to $\Lambda$ in our scheme and so both terms are of order $O(1)$. We can formally write it as
\begin{equation}
    h_B = -P^{AC}_{\ \ B} h_{AC}^{(0)} + \frac{1}{2} \mathring{D}^A h_{AB}^{(-1)} + O\left(\Lambda^{\frac{1}{2}} \right),
\end{equation}
where $P^{AC}_{\ \ \ B}$ is an (non-unique) inverse operator of the one at the left hand side of \eqref{constraint}. Non-uniqueness means that this expression is valid up to the addition of the conformal Killing covector on the round sphere -- we assume this ambiguity to be $\Lambda$-independent (if it had a pole in $\Lambda = 0$, then it would not make any sense to take the limit with our solutions). Analogously, we assume that $\mathcal{E}$ is proportional to $\Lambda^\frac{3}{2}$ as follows from \eqref{weyl}. In this way we derived \eqref{granice}. \\
Let us now use it to estimate different terms in \eqref{horizon_flux}. Since at the horizon $g^{AB} = \frac{\Lambda}{3} \mathring{\gamma}^{AB}$, we have:
\begin{equation}
     h_{2BC} \nabla_D h_{1uA} g^{CD} g^{AB} = O \left(\Lambda^{\frac{3}{2}} \right).
\end{equation}
We also have 
\begin{equation}
    \nabla_r h_{uA} = \partial_r h_{uA} - \Gamma^D_{rA} h_{Du}.
\end{equation}
$\Gamma^D_{rA} = \frac{1}{r} \delta^D_{A} = O(\sqrt{\Lambda})$. Also $\partial_r h_{uA} =  O(\sqrt{\Lambda})$ (as long as \eqref{bondi_exp} is valid). Thus, we can conclude that
\begin{equation}
     h_{2Bu} \nabla_r h_{1uA} g^{AB} = O \left(\Lambda^{\frac{3}{2}} \right).
\end{equation}
The only remaining term is
\begin{equation}
    - \frac{1}{2} h_{2BD} \nabla_u h_{1AC} g^{AB} g^{CD} = - \frac{\Lambda}{6} h^{(-1)}_{2BD} \nabla_u h^{(-1)}_{1AC} \mathring{\gamma}^{AC} \mathring{\gamma}^{BD} + O \left(\Lambda^{\frac{3}{2}} \right),
\end{equation}
which of course reproduces Trautman--Bondi formula (up to the boundary term) as $\Lambda \to 0$.
\section{Momentum} \label{app_mom} \noindent
Calculation of the momenta fluxes requires more care because although 
$$\mathcal{L}_{T_{(i)}} h_{\mu \nu} =: j_{\mu \nu}$$
is a solution to the linearized Einstein equations, it is not written in the Bondi gauge \eqref{bondi_exp}. Indeed, we have:
\begin{align}
    \begin{split}
        j_{AB}^{(0)} &= 2e^{Hu} \mathring{D}_{(A} g h_{B)}^{(0)} \neq 0 \\
        j_{Ar} &= \frac{1}{r^2} e^{Hu} \mathring{D}^B g h_{AB} = \frac{1}{r} e^{Hu} \mathring{D}^B g h_{AB}^{(-1)} + O(r^{-2})  \neq 0 \\
        j_B^{(0)} & = e^{Hu} \left(
        g h_{B,u}^{(0)} - H g h_B^{(0)} - H \mathring{D}^A g \mathring{D}_A h_{B}^{(0)} - H \mathring{D}_B \mathring{D}^C g h_C^{(0)}
        \right) \\
        j_{ur} &=  e^{Hu} \frac{1}{r^2} \mathring{D}^A g h_A = e^{Hu} \mathring{D}^A g h_A^{(0)} + O(r^{-2}) \\
        j_{rr} &= 0. \\
        j_{uu}^{(0)} &= -2H^2 e^{Hu} \mathring{D}^A g_i h_A^{(0)} \label{translacje_pochodna_Lie}
    \end{split} 
\end{align}
In particular, $j_{AB}^{(0)}$ is not traceless with respect to the $\mathring{\gamma}$.
It would be possible to calculate momentum using the very definition. Unfortunately, it is quite messy calculation and so we are in the need of a shortcut. Momenta fluxes were already calculated in \cite{Ashtekar:2015lla}. Moreover, we checked that $Q_T$ and $Q_S$ are the same both in the Bondi gauge and in the Poincar\'e patch. Since we know de Sitter algebra (which we discussed in the Sec. \ref{komutatory}), momenta are already given as a variation of the energy. Since we also checked that this property is satisfied in \cite{Ashtekar:2015lla}, form of the momenta fluxes derived there must necessarily be the same we would obtain. Transforming it conformally, we immediately get 
\begin{equation}
            Q_{T_{(i)}}^P =  \frac{1}{2 H \kappa} \int_{\mathcal{I}^+} d^3 x \sqrt{\mathring{q}}  \mathcal{E}_{cd} \left( \mathcal{L}_{T_{(i)}}H^{-2} h_{ab}^{(0)} - 2e^{Hu} H^{-1} g_{i} h_{ab}^{(0)} \right) \mathring{q}^{ac} \mathring{q}^{bd}.
\end{equation}
In this sense, the second line of \eqref{porownanie} is not a comparison but rather a derivation.
\section{Residual gauge transformations} \noindent \label{residual}
Let us assume that we already have $h_{ab}$ which satisfies Bondi gauge conditions \eqref{bondi_exp} and let us look for the residual gauge transformations generated by $X$. Their action reads
\begin{equation}
    h_{ab} \mapsto h_{ab} + \mathcal{L}_X g_{ab}
\end{equation}
We need to have:
\begin{equation}
    0 =\mathcal{L}_X g_{rr} = -2X^u_{\ ,r}
\end{equation}
so $X^u$ is  $r$-independent. Then, we have
\begin{equation}
    0 =  \mathcal{L}_X g_{rA} = X^B_{\ ,r} g_{AB} - X^u_{\ ,A}
\end{equation}
so
\begin{equation}
    X^B = \mathring{X}^B - \frac{1}{r} \mathring{D}^B X^u.
\end{equation}
From $g_{ru}$ it follows that
\begin{equation}
    X^r = - r X^u_{\ ,u} + \mathring{X}^r.
\end{equation}
Then,
\begin{equation}
    O(r) = \mathcal{L}_X g_{uu} 
\end{equation}
but this is automatically satisfied. The last part is
\begin{equation}
    \mathcal{L}_X g_{ab} = 2r X^r \mathring{\gamma}_{AB} + r^2 \mathcal{L}_{X^B} \mathring{\gamma}_{AB}
\end{equation}
At leading order it is:
\begin{equation}
    -2 X^u_{\ ,u} \mathring{\gamma}_{AB} + \mathring{X}_{(A;B)} = 0
\end{equation}
so it must follows that $\mathring{X}^A$ is a conformal Killing vector on a sphere ($u$-dependent!) and
\begin{equation}
    X^u = \mathring{X}^u  + \frac{1}{2}\int du \mathring{D}_A \mathring{X}^A.
\end{equation}
At the next order we have
\begin{equation}
    2 \mathring{X}^r \mathring{\gamma}_{AB} - 2X^u_{;AB}
\end{equation}
and this is supposed to be traceless, so
\begin{equation}
    \mathring{X}^r = \frac{1}{2} \Delta_{\mathring{\gamma}} X^u
\end{equation}
and the final answer is
\begin{align}
\begin{split}
    X &= \left(\mathring{X}^u + \frac{1}{2}\int du \mathring{D}_A \mathring{X}^A \right) \partial_u + \left(-\frac{1}{2}r \mathring{D}_A \mathring{X}^A + \frac{1}{2} \Delta_{\mathring{\gamma}} \mathring{X}^u + \frac{1}{2}\int du \Delta_{\mathring{\gamma}} \mathring{D}_A \mathring{X}^A  \right) \partial_r \\
    &+ \left(\mathring{X}^B - \frac{1}{r} \mathring{D}^B \mathring{X}^u - \frac{1}{r} \int du \mathring{D}^B \mathring{D}_A \mathring{X}^A \right) \partial_B, \label{res_gen}
\end{split}
\end{align}
where $\mathring{X}^u = \mathring{X}^u(x^A)$ and $\mathring{X}^B$ is $u$-dependent conformal Killing vector on $\mathring{\gamma}$. In particular, all Killing vectors of $g$ are included in this expression. \\
It was proposed \cite{Chrusciel:2020rlz} that one additional condition can be imposed. Intuitively it means that null cones are rigidly transported along $r=0$ line -- there is no $u$-dependent boost or rotation. Technically, it can be stated as follows. At fixed $u$, $h_{uA}^{(0)}$ is a one-form on $\mathbb{S}^2$. From the Hodge--Kodaira decomposition theory it can be expressed as
\begin{equation}
    h_{A}^{(0)} = \mathring{D}_A \hat{\chi} + \mathring{\epsilon}_A^{\ B} \mathring{D}_{B} \check{\chi},
\end{equation}
where $\mathring{\epsilon}$ is the volume form of $\mathring{\gamma}$ and $\hat{\chi}$ and $\check{\chi}$ are $u$-dependent functions on $\mathbb{S}^2$. Then, it was imposed that 
\begin{equation}
    \int_{\mathbb{S}^2} Y_{1m}^\star \hat{\chi} = \int_{\mathbb{S}^2} Y_{1m}^\star \hat{\chi} = 0, \label{rigid_transport}
\end{equation}
where $Y_{1m}$ are arbitrary spherical harmonics with $l=1$. In the $\Lambda = 0$ case, this condition is equivalent to $h_A^{(0)} = 0$ and thus it can be seen as inspired by the asymptotically flat theory. \\
One can check that what $\mathring{X}^B$ does in \eqref{res_gen} is producing linear combinations of $\mathring{D}_A Y_{1m}$ and of $\mathring{\epsilon}_A^{\ B} \mathring{D}_{B} Y_{1m}$. Thus, it must be eliminated unless those linear combinations cancels out. This requirement sets all $\mathring{X}^B$ to zero with the exception of those which corresponds to the Killing vectors of $g$ (since they obviously cannot change $h_{uA}$). Thus, the most general of the residual gauge transformation satisfying rigid transport requirements \eqref{rigid_transport} are
\begin{equation}
    X = f \partial_u + \frac{1}{2}\Delta_{\mathring{\gamma}}f \partial_r - \frac{1}{r} \mathring{D}^B f \partial_B + K^a \partial_a, \label{final res}
\end{equation}
where 
\begin{equation}
    \mathcal{L}_K g = 0
\end{equation}
and $f=f(x^A)$ is any function on a sphere $L^2$-orthogonal to $Y_{1m}$. It is tempting to call a part generated by $f$ supertranslations since they are of the same form as supertranslations in the BMS group. However, it is a misnomer when $\Lambda >0$ because translations are not of this form as can be immediately seen from \eqref{trans}. For the lack of the better name, we propose a little bit different term -- superpseudotranslations.
\bibliography{bibl.bib}

\begin{thebibliography}{27}%
\makeatletter
\providecommand \@ifxundefined [1]{%
 \@ifx{#1\undefined}
}%
\providecommand \@ifnum [1]{%
 \ifnum #1\expandafter \@firstoftwo
 \else \expandafter \@secondoftwo
 \fi
}%
\providecommand \@ifx [1]{%
 \ifx #1\expandafter \@firstoftwo
 \else \expandafter \@secondoftwo
 \fi
}%
\providecommand \natexlab [1]{#1}%
\providecommand \enquote  [1]{``#1''}%
\providecommand \bibnamefont  [1]{#1}%
\providecommand \bibfnamefont [1]{#1}%
\providecommand \citenamefont [1]{#1}%
\providecommand \href@noop [0]{\@secondoftwo}%
\providecommand \href [0]{\begingroup \@sanitize@url \@href}%
\providecommand \@href[1]{\@@startlink{#1}\@@href}%
\providecommand \@@href[1]{\endgroup#1\@@endlink}%
\providecommand \@sanitize@url [0]{\catcode `\\12\catcode `\$12\catcode
  `\&12\catcode `\#12\catcode `\^12\catcode `\_12\catcode `\%12\relax}%
\providecommand \@@startlink[1]{}%
\providecommand \@@endlink[0]{}%
\providecommand \url  [0]{\begingroup\@sanitize@url \@url }%
\providecommand \@url [1]{\endgroup\@href {#1}{\urlprefix }}%
\providecommand \urlprefix  [0]{URL }%
\providecommand \Eprint [0]{\href }%
\providecommand \doibase [0]{http://dx.doi.org/}%
\providecommand \selectlanguage [0]{\@gobble}%
\providecommand \bibinfo  [0]{\@secondoftwo}%
\providecommand \bibfield  [0]{\@secondoftwo}%
\providecommand \translation [1]{[#1]}%
\providecommand \BibitemOpen [0]{}%
\providecommand \bibitemStop [0]{}%
\providecommand \bibitemNoStop [0]{.\EOS\space}%
\providecommand \EOS [0]{\spacefactor3000\relax}%
\providecommand \BibitemShut  [1]{\csname bibitem#1\endcsname}%
\let\auto@bib@innerbib\@empty
\bibitem [{\citenamefont {Kastor}\ and\ \citenamefont
  {Traschen}(2002)}]{kastor2002positive}%
  \BibitemOpen
  \bibfield  {author} {\bibinfo {author} {\bibfnamefont {D.}~\bibnamefont
  {Kastor}}\ and\ \bibinfo {author} {\bibfnamefont {J.}~\bibnamefont
  {Traschen}},\ }\href@noop {} {\bibfield  {journal} {\bibinfo  {journal}
  {Classical and Quantum Gravity}\ }\textbf {\bibinfo {volume} {19}},\ \bibinfo
  {pages} {5901} (\bibinfo {year} {2002})}\BibitemShut {NoStop}%
\bibitem [{\citenamefont {Penrose}(2011)}]{Penrose:2011zza}%
  \BibitemOpen
  \bibfield  {author} {\bibinfo {author} {\bibfnamefont {R.}~\bibnamefont
  {Penrose}},\ }\href {\doibase 10.1007/s10714-011-1255-x} {\bibfield
  {journal} {\bibinfo  {journal} {Gen. Rel. Grav.}\ }\textbf {\bibinfo {volume}
  {43}},\ \bibinfo {pages} {3355} (\bibinfo {year} {2011})}\BibitemShut
  {NoStop}%
\bibitem [{\citenamefont {Ashtekar}\ \emph
  {et~al.}(2015{\natexlab{a}})\citenamefont {Ashtekar}, \citenamefont {Bonga},\
  and\ \citenamefont {Kesavan}}]{Ashtekar:2015lla}%
  \BibitemOpen
  \bibfield  {author} {\bibinfo {author} {\bibfnamefont {A.}~\bibnamefont
  {Ashtekar}}, \bibinfo {author} {\bibfnamefont {B.}~\bibnamefont {Bonga}}, \
  and\ \bibinfo {author} {\bibfnamefont {A.}~\bibnamefont {Kesavan}},\ }\href
  {\doibase 10.1103/PhysRevD.92.044011} {\bibfield  {journal} {\bibinfo
  {journal} {Phys. Rev. D}\ }\textbf {\bibinfo {volume} {92}},\ \bibinfo
  {pages} {044011} (\bibinfo {year} {2015}{\natexlab{a}})},\ \Eprint
  {http://arxiv.org/abs/1506.06152} {arXiv:1506.06152 [gr-qc]} \BibitemShut
  {NoStop}%
\bibitem [{\citenamefont {Szabados}\ and\ \citenamefont
  {Tod}(2015)}]{Szabados:2015wqa}%
  \BibitemOpen
  \bibfield  {author} {\bibinfo {author} {\bibfnamefont {L.~B.}\ \bibnamefont
  {Szabados}}\ and\ \bibinfo {author} {\bibfnamefont {P.}~\bibnamefont {Tod}},\
  }\href {\doibase 10.1088/0264-9381/32/20/205011} {\bibfield  {journal}
  {\bibinfo  {journal} {Class. Quant. Grav.}\ }\textbf {\bibinfo {volume}
  {32}},\ \bibinfo {pages} {205011} (\bibinfo {year} {2015})},\ \Eprint
  {http://arxiv.org/abs/1505.06637} {arXiv:1505.06637 [gr-qc]} \BibitemShut
  {NoStop}%
\bibitem [{\citenamefont {Chruściel}\ \emph {et~al.}(2020)\citenamefont
  {Chruściel}, \citenamefont {Hoque},\ and\ \citenamefont
  {Smołka}}]{Chrusciel:2020rlz}%
  \BibitemOpen
  \bibfield  {author} {\bibinfo {author} {\bibfnamefont {P.~T.}\ \bibnamefont
  {Chruściel}}, \bibinfo {author} {\bibfnamefont {S.~J.}\ \bibnamefont
  {Hoque}}, \ and\ \bibinfo {author} {\bibfnamefont {T.}~\bibnamefont
  {Smołka}},\ }\href@noop {} {\  (\bibinfo {year} {2020})},\ \Eprint
  {http://arxiv.org/abs/2003.09548v2} {arXiv:2003.09548v2 [gr-qc]} \BibitemShut
  {NoStop}%
\bibitem [{\citenamefont {Ashtekar}\ \emph
  {et~al.}(2015{\natexlab{b}})\citenamefont {Ashtekar}, \citenamefont {Bonga},\
  and\ \citenamefont {Kesavan}}]{Ashtekar:2014zfa}%
  \BibitemOpen
  \bibfield  {author} {\bibinfo {author} {\bibfnamefont {A.}~\bibnamefont
  {Ashtekar}}, \bibinfo {author} {\bibfnamefont {B.}~\bibnamefont {Bonga}}, \
  and\ \bibinfo {author} {\bibfnamefont {A.}~\bibnamefont {Kesavan}},\ }\href
  {\doibase 10.1088/0264-9381/32/2/025004} {\bibfield  {journal} {\bibinfo
  {journal} {Class. Quant. Grav.}\ }\textbf {\bibinfo {volume} {32}},\ \bibinfo
  {pages} {025004} (\bibinfo {year} {2015}{\natexlab{b}})},\ \Eprint
  {http://arxiv.org/abs/1409.3816} {arXiv:1409.3816 [gr-qc]} \BibitemShut
  {NoStop}%
\bibitem [{\citenamefont {He}\ and\ \citenamefont {Cao}(2015)}]{He:2015wfa}%
  \BibitemOpen
  \bibfield  {author} {\bibinfo {author} {\bibfnamefont {X.}~\bibnamefont
  {He}}\ and\ \bibinfo {author} {\bibfnamefont {Z.}~\bibnamefont {Cao}},\
  }\href {\doibase 10.1142/S0218271815500819} {\bibfield  {journal} {\bibinfo
  {journal} {Int. J. Mod. Phys. D}\ }\textbf {\bibinfo {volume} {24}},\
  \bibinfo {pages} {1550081} (\bibinfo {year} {2015})}\BibitemShut {NoStop}%
\bibitem [{\citenamefont {Poole}\ \emph {et~al.}(2019)\citenamefont {Poole},
  \citenamefont {Skenderis},\ and\ \citenamefont {Taylor}}]{Poole:2018koa}%
  \BibitemOpen
  \bibfield  {author} {\bibinfo {author} {\bibfnamefont {A.}~\bibnamefont
  {Poole}}, \bibinfo {author} {\bibfnamefont {K.}~\bibnamefont {Skenderis}}, \
  and\ \bibinfo {author} {\bibfnamefont {M.}~\bibnamefont {Taylor}},\ }\href
  {\doibase 10.1088/1361-6382/ab117c} {\bibfield  {journal} {\bibinfo
  {journal} {Class. Quant. Grav.}\ }\textbf {\bibinfo {volume} {36}},\ \bibinfo
  {pages} {095005} (\bibinfo {year} {2019})},\ \Eprint
  {http://arxiv.org/abs/1812.05369} {arXiv:1812.05369 [hep-th]} \BibitemShut
  {NoStop}%
\bibitem [{\citenamefont {Compère}\ \emph {et~al.}(2019)\citenamefont
  {Compère}, \citenamefont {Fiorucci},\ and\ \citenamefont
  {Ruzziconi}}]{Compere:2019bua}%
  \BibitemOpen
  \bibfield  {author} {\bibinfo {author} {\bibfnamefont {G.}~\bibnamefont
  {Compère}}, \bibinfo {author} {\bibfnamefont {A.}~\bibnamefont {Fiorucci}},
  \ and\ \bibinfo {author} {\bibfnamefont {R.}~\bibnamefont {Ruzziconi}},\
  }\href {\doibase 10.1088/1361-6382/ab3d4b} {\bibfield  {journal} {\bibinfo
  {journal} {Class. Quant. Grav.}\ }\textbf {\bibinfo {volume} {36}},\ \bibinfo
  {pages} {195017} (\bibinfo {year} {2019})},\ \Eprint
  {http://arxiv.org/abs/1905.00971} {arXiv:1905.00971 [gr-qc]} \BibitemShut
  {NoStop}%
\bibitem [{\citenamefont {Ashtekar}\ \emph
  {et~al.}(2015{\natexlab{c}})\citenamefont {Ashtekar}, \citenamefont {Bonga},\
  and\ \citenamefont {Kesavan}}]{Ashtekar:2015lxa}%
  \BibitemOpen
  \bibfield  {author} {\bibinfo {author} {\bibfnamefont {A.}~\bibnamefont
  {Ashtekar}}, \bibinfo {author} {\bibfnamefont {B.}~\bibnamefont {Bonga}}, \
  and\ \bibinfo {author} {\bibfnamefont {A.}~\bibnamefont {Kesavan}},\ }\href
  {\doibase 10.1103/PhysRevD.92.104032} {\bibfield  {journal} {\bibinfo
  {journal} {Phys. Rev. D}\ }\textbf {\bibinfo {volume} {92}},\ \bibinfo
  {pages} {104032} (\bibinfo {year} {2015}{\natexlab{c}})},\ \Eprint
  {http://arxiv.org/abs/1510.05593} {arXiv:1510.05593 [gr-qc]} \BibitemShut
  {NoStop}%
\bibitem [{\citenamefont {Bishop}(2016)}]{Bishop:2015kay}%
  \BibitemOpen
  \bibfield  {author} {\bibinfo {author} {\bibfnamefont {N.~T.}\ \bibnamefont
  {Bishop}},\ }\href {\doibase 10.1103/PhysRevD.93.044025} {\bibfield
  {journal} {\bibinfo  {journal} {Phys. Rev. D}\ }\textbf {\bibinfo {volume}
  {93}},\ \bibinfo {pages} {044025} (\bibinfo {year} {2016})},\ \Eprint
  {http://arxiv.org/abs/1512.05663} {arXiv:1512.05663 [gr-qc]} \BibitemShut
  {NoStop}%
\bibitem [{\citenamefont {Date}\ and\ \citenamefont
  {Hoque}(2016)}]{Date:2015kma}%
  \BibitemOpen
  \bibfield  {author} {\bibinfo {author} {\bibfnamefont {G.}~\bibnamefont
  {Date}}\ and\ \bibinfo {author} {\bibfnamefont {S.~J.}\ \bibnamefont
  {Hoque}},\ }\href {\doibase 10.1103/PhysRevD.94.064039} {\bibfield  {journal}
  {\bibinfo  {journal} {Phys. Rev. D}\ }\textbf {\bibinfo {volume} {94}},\
  \bibinfo {pages} {064039} (\bibinfo {year} {2016})},\ \Eprint
  {http://arxiv.org/abs/1510.07856} {arXiv:1510.07856 [gr-qc]} \BibitemShut
  {NoStop}%
\bibitem [{\citenamefont {Abbott}\ \emph {et~al.}(2016)\citenamefont {Abbott}
  \emph {et~al.}}]{Abbott:2016blz}%
  \BibitemOpen
  \bibfield  {author} {\bibinfo {author} {\bibfnamefont {B.}~\bibnamefont
  {Abbott}} \emph {et~al.} (\bibinfo {collaboration} {LIGO Scientific,
  Virgo}),\ }\href {\doibase 10.1103/PhysRevLett.116.061102} {\bibfield
  {journal} {\bibinfo  {journal} {Phys. Rev. Lett.}\ }\textbf {\bibinfo
  {volume} {116}},\ \bibinfo {pages} {061102} (\bibinfo {year} {2016})},\
  \Eprint {http://arxiv.org/abs/1602.03837} {arXiv:1602.03837 [gr-qc]}
  \BibitemShut {NoStop}%
\bibitem [{\citenamefont {Riess}\ \emph {et~al.}(1998)\citenamefont {Riess},
  \citenamefont {Filippenko}, \citenamefont {Challis}, \citenamefont
  {Clocchiatti}, \citenamefont {Diercks}, \citenamefont {Garnavich},
  \citenamefont {Gilliland}, \citenamefont {Hogan}, \citenamefont {Jha},
  \citenamefont {Kirshner} \emph {et~al.}}]{riess1998observational}%
  \BibitemOpen
  \bibfield  {author} {\bibinfo {author} {\bibfnamefont {A.~G.}\ \bibnamefont
  {Riess}}, \bibinfo {author} {\bibfnamefont {A.~V.}\ \bibnamefont
  {Filippenko}}, \bibinfo {author} {\bibfnamefont {P.}~\bibnamefont {Challis}},
  \bibinfo {author} {\bibfnamefont {A.}~\bibnamefont {Clocchiatti}}, \bibinfo
  {author} {\bibfnamefont {A.}~\bibnamefont {Diercks}}, \bibinfo {author}
  {\bibfnamefont {P.~M.}\ \bibnamefont {Garnavich}}, \bibinfo {author}
  {\bibfnamefont {R.~L.}\ \bibnamefont {Gilliland}}, \bibinfo {author}
  {\bibfnamefont {C.~J.}\ \bibnamefont {Hogan}}, \bibinfo {author}
  {\bibfnamefont {S.}~\bibnamefont {Jha}}, \bibinfo {author} {\bibfnamefont
  {R.~P.}\ \bibnamefont {Kirshner}},  \emph {et~al.},\ }\href@noop {}
  {\bibfield  {journal} {\bibinfo  {journal} {The Astronomical Journal}\
  }\textbf {\bibinfo {volume} {116}},\ \bibinfo {pages} {1009} (\bibinfo {year}
  {1998})}\BibitemShut {NoStop}%
\bibitem [{\citenamefont {Chru\'sciel}\ and\ \citenamefont
  {Ifsits}(2016)}]{Chrusciel:2016oux}%
  \BibitemOpen
  \bibfield  {author} {\bibinfo {author} {\bibfnamefont {P.~T.}\ \bibnamefont
  {Chru\'sciel}}\ and\ \bibinfo {author} {\bibfnamefont {L.}~\bibnamefont
  {Ifsits}},\ }\href {\doibase 10.1103/PhysRevD.93.124075} {\bibfield
  {journal} {\bibinfo  {journal} {Phys. Rev. D}\ }\textbf {\bibinfo {volume}
  {93}},\ \bibinfo {pages} {124075} (\bibinfo {year} {2016})},\ \Eprint
  {http://arxiv.org/abs/1603.07018} {arXiv:1603.07018 [gr-qc]} \BibitemShut
  {NoStop}%
\bibitem [{\citenamefont {Friedrich}(1986)}]{friedrich1986}%
  \BibitemOpen
  \bibfield  {author} {\bibinfo {author} {\bibfnamefont {H.}~\bibnamefont
  {Friedrich}},\ }\href {https://projecteuclid.org:443/euclid.cmp/1104116232}
  {\bibfield  {journal} {\bibinfo  {journal} {Comm. Math. Phys.}\ }\textbf
  {\bibinfo {volume} {107}},\ \bibinfo {pages} {587} (\bibinfo {year}
  {1986})}\BibitemShut {NoStop}%
\bibitem [{\citenamefont {Ashtekar}\ \emph {et~al.}(2016)\citenamefont
  {Ashtekar}, \citenamefont {Bonga},\ and\ \citenamefont
  {Kesavan}}]{Ashtekar:2015ooa}%
  \BibitemOpen
  \bibfield  {author} {\bibinfo {author} {\bibfnamefont {A.}~\bibnamefont
  {Ashtekar}}, \bibinfo {author} {\bibfnamefont {B.}~\bibnamefont {Bonga}}, \
  and\ \bibinfo {author} {\bibfnamefont {A.}~\bibnamefont {Kesavan}},\ }\href
  {\doibase 10.1103/PhysRevLett.116.051101} {\bibfield  {journal} {\bibinfo
  {journal} {Phys. Rev. Lett.}\ }\textbf {\bibinfo {volume} {116}},\ \bibinfo
  {pages} {051101} (\bibinfo {year} {2016})},\ \Eprint
  {http://arxiv.org/abs/1510.04990} {arXiv:1510.04990 [gr-qc]} \BibitemShut
  {NoStop}%
\bibitem [{\citenamefont {Trautman}(1958)}]{Trautman:2016xic}%
  \BibitemOpen
  \bibfield  {author} {\bibinfo {author} {\bibfnamefont {A.}~\bibnamefont
  {Trautman}},\ }\href@noop {} {\bibfield  {journal} {\bibinfo  {journal}
  {Bull. Acad. Pol. Sci. Ser. Sci. Math. Astron. Phys.}\ }\textbf {\bibinfo
  {volume} {6}},\ \bibinfo {pages} {407} (\bibinfo {year} {1958})},\ \Eprint
  {http://arxiv.org/abs/1604.03145} {arXiv:1604.03145 [gr-qc]} \BibitemShut
  {NoStop}%
\bibitem [{\citenamefont {Bondi}\ \emph {et~al.}(1962)\citenamefont {Bondi},
  \citenamefont {van~der Burg},\ and\ \citenamefont {Metzner}}]{Bondi:1962px}%
  \BibitemOpen
  \bibfield  {author} {\bibinfo {author} {\bibfnamefont {H.}~\bibnamefont
  {Bondi}}, \bibinfo {author} {\bibfnamefont {M.}~\bibnamefont {van~der Burg}},
  \ and\ \bibinfo {author} {\bibfnamefont {A.}~\bibnamefont {Metzner}},\ }\href
  {\doibase 10.1098/rspa.1962.0161} {\bibfield  {journal} {\bibinfo  {journal}
  {Proc. Roy. Soc. Lond. A}\ }\textbf {\bibinfo {volume} {A269}},\ \bibinfo
  {pages} {21} (\bibinfo {year} {1962})}\BibitemShut {NoStop}%
\bibitem [{\citenamefont {Sachs}(1962)}]{Sachs:1962wk}%
  \BibitemOpen
  \bibfield  {author} {\bibinfo {author} {\bibfnamefont {R.}~\bibnamefont
  {Sachs}},\ }\href {\doibase 10.1098/rspa.1962.0206} {\bibfield  {journal}
  {\bibinfo  {journal} {Proc. Roy. Soc. Lond. A}\ }\textbf {\bibinfo {volume}
  {A270}},\ \bibinfo {pages} {103} (\bibinfo {year} {1962})}\BibitemShut
  {NoStop}%
\bibitem [{\citenamefont {Christodoulou}\ and\ \citenamefont
  {Klainerman}(1993)}]{christodoulou1993global}%
  \BibitemOpen
  \bibfield  {author} {\bibinfo {author} {\bibfnamefont {D.}~\bibnamefont
  {Christodoulou}}\ and\ \bibinfo {author} {\bibfnamefont {S.}~\bibnamefont
  {Klainerman}},\ }\href@noop {} {\bibfield  {journal} {\bibinfo  {journal}
  {S{\'e}minaire {\'E}quations aux d{\'e}riv{\'e}es partielles
  (Polytechnique)}\ ,\ \bibinfo {pages} {1}} (\bibinfo {year}
  {1993})}\BibitemShut {NoStop}%
\bibitem [{\citenamefont {Moncrief}(1976)}]{Moncrief:1976un}%
  \BibitemOpen
  \bibfield  {author} {\bibinfo {author} {\bibfnamefont {V.}~\bibnamefont
  {Moncrief}},\ }\href {\doibase 10.1063/1.522814} {\bibfield  {journal}
  {\bibinfo  {journal} {J. Math. Phys.}\ }\textbf {\bibinfo {volume} {17}},\
  \bibinfo {pages} {1893} (\bibinfo {year} {1976})}\BibitemShut {NoStop}%
\bibitem [{\citenamefont {Penrose}(1965)}]{penrose1965zero}%
  \BibitemOpen
  \bibfield  {author} {\bibinfo {author} {\bibfnamefont {R.}~\bibnamefont
  {Penrose}},\ }\href@noop {} {\bibfield  {journal} {\bibinfo  {journal}
  {Proceedings of the Royal Society of London. Series A. Mathematical and
  physical sciences}\ }\textbf {\bibinfo {volume} {284}},\ \bibinfo {pages}
  {159} (\bibinfo {year} {1965})}\BibitemShut {NoStop}%
\bibitem [{\citenamefont {Ashtekar}\ and\ \citenamefont
  {Bahrami}(2019)}]{Ashtekar:2019khv}%
  \BibitemOpen
  \bibfield  {author} {\bibinfo {author} {\bibfnamefont {A.}~\bibnamefont
  {Ashtekar}}\ and\ \bibinfo {author} {\bibfnamefont {S.}~\bibnamefont
  {Bahrami}},\ }\href {\doibase 10.1103/PhysRevD.100.024042} {\bibfield
  {journal} {\bibinfo  {journal} {Phys. Rev. D}\ }\textbf {\bibinfo {volume}
  {100}},\ \bibinfo {pages} {024042} (\bibinfo {year} {2019})},\ \Eprint
  {http://arxiv.org/abs/1904.02822} {arXiv:1904.02822 [gr-qc]} \BibitemShut
  {NoStop}%
\bibitem [{\citenamefont {Date}\ and\ \citenamefont
  {Hoque}(2017)}]{Date:2016uzr}%
  \BibitemOpen
  \bibfield  {author} {\bibinfo {author} {\bibfnamefont {G.}~\bibnamefont
  {Date}}\ and\ \bibinfo {author} {\bibfnamefont {S.~J.}\ \bibnamefont
  {Hoque}},\ }\href {\doibase 10.1103/PhysRevD.96.044026} {\bibfield  {journal}
  {\bibinfo  {journal} {Phys. Rev. D}\ }\textbf {\bibinfo {volume} {96}},\
  \bibinfo {pages} {044026} (\bibinfo {year} {2017})},\ \Eprint
  {http://arxiv.org/abs/1612.09511} {arXiv:1612.09511 [gr-qc]} \BibitemShut
  {NoStop}%
\bibitem [{\citenamefont {Hoque}\ and\ \citenamefont
  {Virmani}(2018)}]{Hoque:2018byx}%
  \BibitemOpen
  \bibfield  {author} {\bibinfo {author} {\bibfnamefont {S.~J.}\ \bibnamefont
  {Hoque}}\ and\ \bibinfo {author} {\bibfnamefont {A.}~\bibnamefont
  {Virmani}},\ }\href {\doibase 10.1007/s10714-018-2359-3} {\bibfield
  {journal} {\bibinfo  {journal} {Gen. Rel. Grav.}\ }\textbf {\bibinfo {volume}
  {50}},\ \bibinfo {pages} {40} (\bibinfo {year} {2018})},\ \Eprint
  {http://arxiv.org/abs/1801.05640} {arXiv:1801.05640 [gr-qc]} \BibitemShut
  {NoStop}%
\bibitem [{\citenamefont {Hollands}\ and\ \citenamefont
  {Wald}(2013)}]{Hollands:2012sf}%
  \BibitemOpen
  \bibfield  {author} {\bibinfo {author} {\bibfnamefont {S.}~\bibnamefont
  {Hollands}}\ and\ \bibinfo {author} {\bibfnamefont {R.~M.}\ \bibnamefont
  {Wald}},\ }\href {\doibase 10.1007/s00220-012-1638-1} {\bibfield  {journal}
  {\bibinfo  {journal} {Commun. Math. Phys.}\ }\textbf {\bibinfo {volume}
  {321}},\ \bibinfo {pages} {629} (\bibinfo {year} {2013})},\ \Eprint
  {http://arxiv.org/abs/1201.0463} {arXiv:1201.0463 [gr-qc]} \BibitemShut
  {NoStop}%
\end{thebibliography}%
\end{document}